\documentclass[10pt]{iopart}
\usepackage{iopams}
\usepackage{graphicx,epstopdf}
\usepackage{braket}
\usepackage{color}
\usepackage{float}
\usepackage{epsfig}
\usepackage{epstopdf}
\usepackage{times}
\usepackage{cite}
\usepackage[normalem]{ulem}

\usepackage[colorlinks=true, citecolor=blue, urlcolor=blue ]{hyperref}
\begin{document}

\title{Self-bound Bose-Fermi liquids in lower dimensions}

\author{Debraj Rakshit,$\,^{1,2}$ Tomasz Karpiuk,$\,^3$ Pawe{\l} Zin,$\,^4$ Miros{\l}aw Brewczyk,$\,^3$ Maciej Lewenstein$\,^{2,5}$ and Mariusz Gajda$\,^1$}

\address{
\mbox{$^1$ Institute of Physics, Polish Academy of Sciences, Aleja Lotnik{\'o}w 32/46, PL-02668 Warsaw, Poland}
{$^2$ ICFO - Institut de Ciencies Fotoniques, The Barcelona Institute of Science and Technology, Av. Carl Friedrich Gauss 3, 08860 Castelldefels (Barcelona), Spain} \\
\mbox{$^3$ Wydzia{\l} Fizyki, Uniwersytet w Bia{\l}ymstoku,  ul. K. Cio{\l}kowskiego 1L, 15-245 Bia{\l}ystok, Poland} \\
\mbox{$^4$ National Centre for Nuclear Research, ul. Ho\.za 69, PL-00-681 Warsaw, Poland} \\
\mbox{$^5$ ICREA, Pg. Lluis Companys 23, 08010 Barcelona, Spain}
}

\begin{abstract}
We study weakly interacting mixtures of ultracold atoms composed of bosonic and fermionic species  in 2D and 1D. When interactions between particles are appropriately tuned, self-bound quantum liquids can be formed.  We show that while formation of these droplets  in 2D is due to the higher order correction terms contributing to the total energy and originating in quantum fluctuations, in 1D geometry the quantum fluctuations have a negligible role on formation of the self-bound systems. The leading mean-field interactions are then sufficient for droplet formation in 1D.  
We analyse stability conditions for 2D and 1D systems and predict values of equilibrium densities of droplets.

\end{abstract}


\submitto{\NJP}

\noindent
\section{Introduction}
\label{intro}
Although the physical space is three-dimensional, lower dimensional spaces are quite often considered on a theoretical ground. This is because they are simpler for analysis,  nonetheless  able to capture the basic physics of the three-dimensional space. On the other hand the Hall effect, fractional statistics, Berezinskii-Kosterlitz-Thoules transition, Tonks-Girardeau gas are only some examples of physical phenomena attributed exclusively to 2D or 1D spaces.

It is possible to impose constraints on real 3D systems such that from the kinematic point of view they behave as low dimensional. In a case of ultracold quantum gases such tight confinement in one or two spacial directions can be obtained by a proper shaping of external fields forming a trap, making them  highly anisotropic. If the excitation energy in the tightly confined direction(s) is the largest energy scale of the problem, the lower dimensional physics comes into a play.

Recently quantum liquids attracted a lot of attention of both experimentalists \cite{Kadau16, Ferrier16a, Schmitt16, Chomaz16, Cabrera17, Tarruell17b, Fattori17} and theorists \cite{Petrov15,Wachtler16a,Baillie16,Rafal16,Cui18}. Quantum self bound droplets can be  formed in dipolar condensates or two-component Bose-Bose mixtures \cite{Petrov15}. Recently conditions for formation of droplets in a Bose-Fermi mixture were found \cite{Rakshit18}. The life-time of droplets is limited by three-body collisions and at present experiments it is of the order of tens milliseconds. In lower dimensions it is expected that this life-time can be extended because of reduced phase-space available to colliding atoms. Low dimensional systems are therefore of the key interest \cite{Petrov16,Mishra16,Zin18,Ilg18}. In this work we study a possibility of formation of the self-bound low dimensional liquid droplets in a two component mixture of ultracold bosonic and fermionic atoms.\\

\section{Uniform 2D mixture}
\label{uni2d}
The mean-field energy density of an uniform Bose-Fermi mixture is given by
\begin{equation}
\varepsilon_{MF} = \frac{1}{2}\beta n_F^2  + \frac{1}{2} g_{BB} n_B^2 + g_{BF} n_B n_F,
\label{emf}
\end{equation}
where $n_B$ and $n_F$ are bosonic and fermionic densities, respectively, and $\beta=2 \pi \hbar^2/m_F$. The first term in Eq.~(\ref{emf}) corresponds to kinetic energy of fermions. The next two terms account for  energy of intra-species (boson-boson) and inter-species (boson-fermion) interactions. The coupling constant $g_{\sigma \sigma'}$ is related to the 2D scattering lengths $a_{\sigma \sigma'}$ by $g_{\sigma \sigma'}=\frac{2 \pi \hbar^2}{\mu_{\sigma \sigma'}}\frac{1}{ \ln(\epsilon/(a_{\sigma \sigma'}^2 \kappa^2))}$, where $\kappa$ is a cut-off momentum, $\epsilon=4\exp(-2\gamma)$, $\gamma$ is Euler's constant, and $\mu_{\sigma \sigma'}=m_{\sigma}m_{\sigma'}/(m_{\sigma}+m_{\sigma}')$ is the reduced mass \cite{Petrov16,Popov71}. Here $\sigma  \in \{B,F\}$. 

2D scattering has few key differences in comparison to the 3D case -- the 2D scattering length, $a_{2d}$, is always positive valued; scattering amplitude is energy dependent for all possible values of $a_{BB}$ and $a_{BF}$. The dependence of energy is weak, so it is always possible to use a fixed value of coupling strength provided that high energy contributions are eliminated by a proper choice of a value of a cut-off momentum $\kappa$.  The weakly-interacting regime requires $a_{\sigma \sigma'} \ll r_0$ (weak repulsion) or $a_{\sigma \sigma'} \gg r_0$ (weak attraction). Here $r_0$ is inter-particle separation. We show that it is possible to find such energy range of colliding particles, and thus to choose a suitable value for the cut-off momentum, for which the intra-species interaction is weakly repulsive and interspecies interaction is weakly attractive.

In the case of a two-dimensional Bose gas we deal with a quasicondensate. In such a case the lowest order correction to the mean field energy is given by \cite{Castin03}  
\begin{equation}
\varepsilon_{LHY}=E_{LHY}/\Omega=\frac{1}{2} \sum_{k<\kappa} \left[ \omega_k - g_{BB} n_B - \frac{\hbar^2 k^2}{2 m_B}\right],
\label{elhy}
\end{equation}
where $\varepsilon_{LHY}$ is the Lee-Huang-Yang (LHY) \cite{Lee57} energy density, $\Omega$ is a volume entity, $\omega_k=(g_{BB} n_B \hbar^2 k^2/m_B + (\hbar^2 k^2/2 m_B)^2)^{1/2}$. The contribution to the bosonic repulsive energy, $E_{LHY}=\Omega \varepsilon_{LHY}$, due to the zero-point energy can be obtained via standard Bogoliubov theory.

The above expression, representing interacting Bose gas, was obtained  by treating ultraviolet divergences using  the discreet lattice model as a regularization scheme. Therefore integration in Eq.~(\ref{elhy}) over momentum is carried up to the cut-off momentum $\kappa$. The  2D integration  gives the following contribution: 
\begin{equation}
\varepsilon_{LHY}=\frac{g_{BB}^2 n_B^2 m_B}{8 \pi \hbar^2}  \ln\left(\frac{g_{BB} n_B \sqrt{e}}{\hbar^2\kappa^2/m_B}\right).
\label{elhy1}
\end{equation}

The higher-order correction in the Bose-Fermi coupling in a three-dimensional mixture were estimated via several theoretical approaches \cite{Wilkens02,Viverit02,Zhai11}, the results from which agree within appropriate validity regimes. In order to evaluate the correction to the Bose-Fermi interaction due to coupling between density fluctuations in the two species, we begin with the relevant term given in Ref.~\cite{Viverit02}:
\begin{equation}
\delta \varepsilon_{BF}= \frac{-n_B g_{BF}^2}{\Omega^2}\sum_{|\vec{k}|<\kappa,|\vec{q}|<\kappa} (u_k+v_k)^2 \frac{n_{\vec{q}}^F (1-n_{\vec{q}+\vec{k}}^F)}{\omega_k+\epsilon_{|\vec{q}+\vec{k}|}^F-\epsilon_q^F}.
\label{bf-correction}
\end{equation}
Similarly as for Bose system we use here a discreet lattice model as the  regularization scheme, and thus integration over momenta are carried up to the cut-off momentum determined by the lattice. Therefore the term used in \cite{Viverit02} that counters ultraviolet divergence has been dropped. Here $\epsilon_k^{B,F}=\frac{\hbar^2 k^2}{2 m_{B,F}}$, $\omega_k=\left( (\epsilon_k^B)^2+2 g_{BB} n_B \epsilon_k^B \right)^{1/2}$ and $n_q^F=\Theta(k_F-q)$. The Fermi momentum is given by $k_F=\sqrt{4 \pi n_F}$. The coefficients $u_k$ and $v_k$ follow from the relations: $u_k^2=1+v_k^2=(\epsilon_k^B+g_{BB} n_B+\omega_k)/2 \omega_k$ and $2u_k v_k=-g_{BB} n_B/2\omega_k$. Finally, we obtain
\begin{eqnarray}
\delta \varepsilon_{BF}=-\frac{2 \pi n_B g_{BF}^2 2 m_B k_F^2}{(2 \pi)^4 \hbar^2} \mathcal{I}_c,
\end{eqnarray}
where
\begin{eqnarray}
\mathcal{I}_c = \int_0^{\tilde{k}_c} d\tilde{k} \int_0^1 d\tilde{q} \int_0^{2\pi} d\theta
\nonumber\\
\frac{\tilde{k}\tilde{q}\left(1-\Theta \left(1-\sqrt{\tilde{k}^2+\tilde{q}^2+2\tilde{q}\tilde{k}\cos \theta}\right)\right)}{\sqrt{\tilde{k}^2+\alpha} \left( \sqrt{\tilde{k}^2+\alpha} + w \tilde{k}+2 w \tilde{q} \cos\theta\right)}.
\end{eqnarray}
Here $\tilde{k}_c=\kappa/k_F$ and $\alpha=2 w (g_{BB}n_B/\epsilon_F)$, where $\epsilon_F$ is the Fermi energy, and $w=m_B/m_F$.

\begin{figure}[t]
\begin{center}
\includegraphics[width=6.5cm,height=8.5cm]{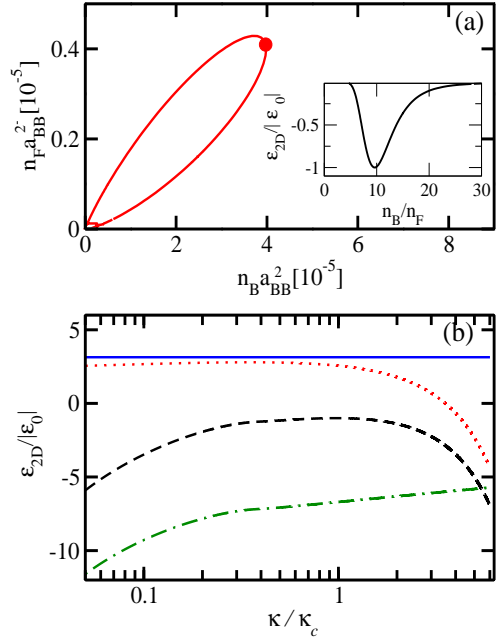}
\caption{2D case of $^{133}$Cs-$^{6}$Li mixture with $a_{BF}/a_{BB}=10^4$. (a) Red solid line shows the zero-pressure contour. The equilibrium density is marked by the red dot. Inset: Energy density, $\varepsilon_{2D}$, as a function of $n_B/n_F$, subjected to the zero-pressure constraint. The vertical axis is rescaled by the equilibrium energy density, $|\varepsilon_0|$. The minimum corresponds to the red dot on the zero-pressure contour. (b) Variation of various contribution to the total energy density as a function of the cut-off momentum, $\kappa$, around the chosen cut-off momentum $\kappa_c$ obtained via iterative procedure. The blue solid line shows fermionic kinetic energy density. The red dotted line and the green dashed-dotted line correspond to $\varepsilon_B$ and $\varepsilon_{BF}$, respectively. The total energy density of the system (Eq.~(\ref{e2d})) is marked by black dashed line.}
\label{fig1}
\end{center}
\end{figure}

Let $\varepsilon_B=g_{BB}n_B^2/2+\varepsilon_{LHY}$ and  $\varepsilon_{BF}=g_{BF} n_B n_F+\delta \varepsilon_{BF}$. The total energy density of the system:
\begin{equation}
\varepsilon_{2D} = \beta n_F^2/2+\varepsilon_B+\varepsilon_{BF}
\label{e2d}
\end{equation}
formally depends on a cut-off momentum $\kappa$ via coupling constants and to specify the system the value of $\kappa=\kappa_c$ has to be set. If chosen correctly  the total energy density of the system is almost independent of the choice. Guided  by analogous analysis of D. Petrov and G. Astrakharchik \cite{Petrov16} who considered Bose-Bose droplet in two dimensions, we observe that the energy density reaches a maximal value when varying the cut-off at fixed values of bosonic and fermionic densities, $\partial \varepsilon_{2D}/\partial {\kappa}^2|_{\kappa=\kappa_c}=0$. $\kappa_c$ is our choice for the cut-off momentum. Evidently, it  depends on the densities of the species. Two comments are rather obvious: i) close to the maximum there is no dependence on $\kappa$, ii) for correctly chosen cut-off, the beyond-mean field terms should be small corrections to the leading  mean-field energy.  
 
To better justify this choice for the Bose-Fermi system we consider a limit of infinitely weak attraction between the two components of the mixture, i.e. $n_{B}a_{BF}^2 \rightarrow \infty$. In this limit bosonic and fermionic atoms become independent. Energy of bosonic subsystem in the limit of vanishing density, $n_B \rightarrow 0$, should approach the energy of a two-dimensional single component Bose 
gas, given by the famous Schick formula \cite{Schick71}. In  \ref{justification} we show that this is indeed the case. This is  a very important suggestion that our choice of the cut-off  shall give correct value of ground state energy of Bose-Fermi system.

The cut-off depends on densities, but on the other hand, the densities of species forming a droplet are fixed by interactions, thus depend (weakly) on the cut-off. The problem of choosing the cut-off must be solved in a self-consistent way. Our strategy is the following. If the cut-off is set then the interactions' couplings are fixed. Stability conditions of the self-bound system give  equilibrium densities of a droplet.  Having those, the cut-off corresponding the maximum of  energy, for fixed atomic densities,  can be determined. The self-consistency can be easily  reached. The procedure converges very fast and gives both, the optimal cut-off and the densities of Bose and Fermi components.


We focus here at the weakly interacting regime  assuming Bose-Fermi attraction, i.e. $a_{BF}^2 n_{B,F} \gg 1$, and Bose-Bose repulsion, i.e. $a_{BB}^2 n_{B,F} \ll 1$. The initial guess for the cut-off momentum is  chosen from the bare minimum stability condition of the mixture, similarly as in  \cite{Petrov16,monte}. Determinant of the second derivatives of mean-field energies must vanish at the stability edge: $(\partial^2\varepsilon_{MF}/\partial n_B^2)(\partial^2\varepsilon_{MF}/\partial n_F^2)|_{\kappa=\kappa_c}-(\partial^2\varepsilon_{MF}/\partial n_B \partial n_F)^2|_{\kappa=\kappa_c}=0$. This leads us to the following equation:
\begin{eqnarray}
2 \ln^2\left(\frac{\epsilon}{a_{BF}^2 \kappa_c^2}\right)=\frac{(1+w)^2}{w} \ln\left(\frac{\epsilon}{a_{BB}^2\kappa_c^2}\right),
\label{cut-off}
\end{eqnarray}
which can be solved for $\kappa_c$, and the abovementioned iterative procedure can be initiated. The initial value of $\kappa_c$ happens to be a quite good guess.\\

The necessary condition for the appearance of a liquid droplet is vanishing pressure i.e.,
\begin{equation}
p = \varepsilon_{2D} - \mu_F n_F - \mu_B n_B =0,
\label{zero-pressure}
\end{equation}
where the chemical potentials $\mu_{\sigma}=\partial \varepsilon_{2D}/\partial n_{\sigma}$. Due to the quadratic form of the dominant mean-field energy of the Bose-Fermi mixture in 2D, at the edge of the mean-field instability, i.e., for a small and negative $\delta g = g_{BF}+\sqrt{\beta g_{BB}}$, we can use arguments of \cite{Petrov15} and show that the equilibrium densities are close to the line $n_B^{eq}/n_F^{eq} \approx \sqrt{\beta/g_{BB}}$, which turns out to be
\begin{equation}
{n_B^{eq}}/{n_F^{eq}} \approx \left(\frac{w \ln (\epsilon/(a_{BB}^2 \kappa_c^2))}{2}\right)^{1/2}.
\label{eq_ratio}
\end{equation}

In the following we consider a weakly interacting mixture of $^{133}$Cs and $^{6}$Li, corresponding to the mass ratio $w = 22.09$, with varied $a_{BF}/a_{BB}$.
Fig.~\ref{fig1}(a) demonstrates a graphical representation of the numerical solution from Eq.~(\ref{zero-pressure}) for a particular case of $^{133}$Cs-$^{6}$Li with $a_{BF} /a_{BB} = 10^4$. The zero-pressure line forms a closed contour on $n_B-n_F$ plane.  The droplet forms at equilibrium densities, for which the energy density of the system constrained by the zero-pressure requirement (Eq.~(\ref{zero-pressure})) attains the minimal value. This implies
\begin{equation}
\mu_B \frac{\partial p}{\partial n_F} - \mu_F \frac{\partial p}{\partial n_B} = 0.
\label{zero-pressure-2}
\end{equation}
In Fig.~\ref{fig1}(a), the energy minimum is marked by a dot. The variation of the energy density, $\varepsilon_{2D}$, along the zero-pressure line is shown in the inset of Fig.~\ref{fig1}(a). The particular density ratio $n_B/n_F$, for which the system's total energy density is minimal, corresponds to the dot at the zero-pressure contour.

In Fig.~\ref{fig1}(b), we examine the flatness of various components of the total energy density functional around the chosen cut-off momentum, $a_B \kappa_c=0.034$ obtained via the self-consistent approach. One can see that close to $\kappa_c$, once the higher-order corrections are accounted for along with the mean-field contributions, the energy density functionals show only a slight variance with the cut-off momentum. This suggests existence of a plateau around $\kappa=\kappa_c$, where droplet equilibrium densities are almost independent on the choice of a particular cut-off.

\begin{figure}[t]
\begin{center}
\includegraphics[width=6.6cm,height=4.5cm]{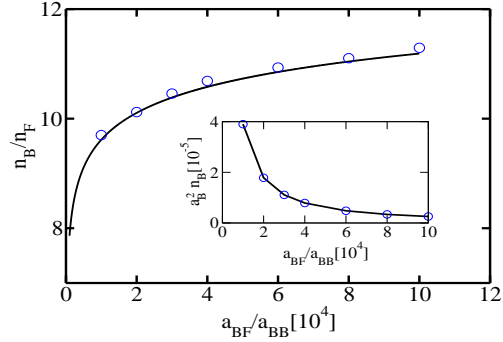}
\caption{Bosonic to fermionic density ratio of $^{133}$Cs-$^{6}$Li mixture at equilibrium as a function of $a_{BF}/a_{BB}$. The solid line is the analytic result obtained from Eq.~(\ref{eq_ratio}). The red circles represent the same obtained from complete analysis based on Eqs.~(\ref{zero-pressure}) and (\ref{zero-pressure-2}). Inset: The blue circles  show equilibrium density of the bosonic species, Cs, as a function of $a_{BF}/a_{BB}$.}
\label{fig2}
\end{center}
\end{figure}

In Fig.~\ref{fig2}, we show the equilibrium density ratios of liquid droplets in $^{133}$Cs-$^{6}$Li mixture with $a_{BF}/a_{BB}$. The solid line shows the approximate results governed by Eq.~(\ref{eq_ratio}), and the circles represent the full solutions obtained via direct inspections of energy minima on the zero-pressure contour as explained in Fig.~\ref{fig1}.  The agreement between the complete and approximate solutions remains very good. In the inset of Fig.~\ref{fig2}, we show the bosonic equilibrium density as a function of $a_{BF}/a_{BB}$. The fermionic part behaves similarly.

\section{Finite 2D mixture} 
\label{finite2d}

\begin{figure}[t]
\begin{center}
\includegraphics[width=6.6cm,height=4.5cm]{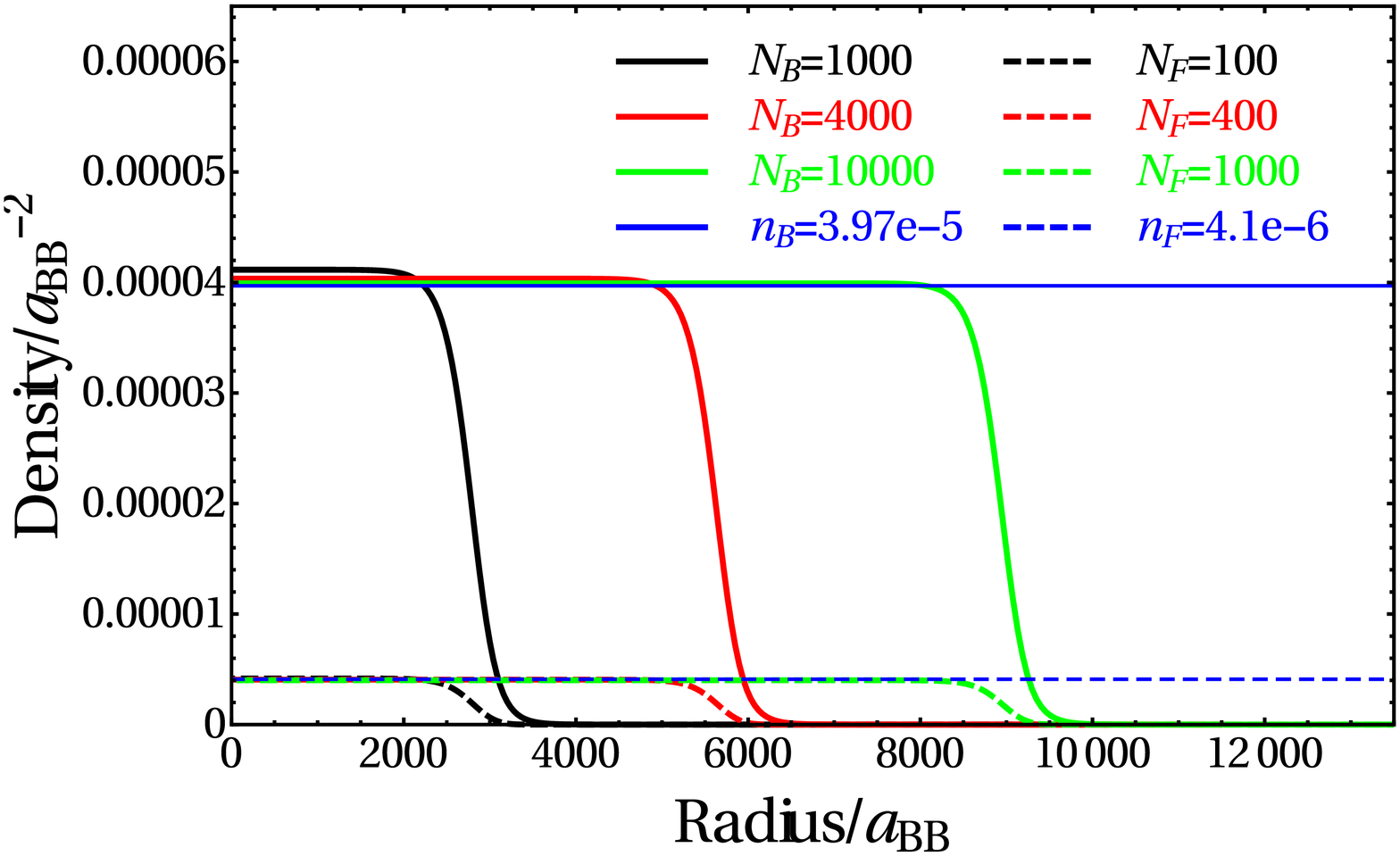}
\caption{Radial densities (solid and dashed lines for bosons and fermions, respectively) of the Bose-Fermi droplets for $^{133}$Cs$\;$-$^{6}$Li mixture for  $a_{BF}/a_{BB}=10^4$ and the initial number of bosons (fermions) equal to $1000$ ($100$), $4000$ ($400$), and $10000$ ($1000$).  The horizontal lines are the bosonic and fermionic densities coming from the analysis ignoring the surface effects. Clearly, the surface effects for larger droplets can be neglected.}
\label{fig3}
\end{center}
\end{figure}

Now we investigate the density profile of the Bose-Fermi droplets with finite number of particles. We follow the similar treatment adopted in context of the 3D Bose-Fermi droplets \cite{Rakshit18}. In order to incorporate the surface effects, it's required to consider additional density gradient terms in system energy. This is done within local density approximation via inclusion of the kinetic energy $E_k^B = \int d\mathbf{r}\, \varepsilon_k^B$ with  $\varepsilon_k^B = (\hbar^2/{2m_B}) ({\nabla} \sqrt{n_B})^2$ for bosonic component and the  Weizs{\"a}cker correction \cite{Weizsacker} to the kinetic energy, $E_{k,W}^F = \int d\mathbf{r}\, \varepsilon_{k,W}^F$, where $\varepsilon_{k,W}^F = \xi\, (\hbar^2/8m_F)\, (\nabla n_F)^2/n_F$, for the fermionic component. For a 3D Fermi gas $\xi=1/9$ \cite{Kirznits, Oliver}, for a 2D case (Eq.~(\ref{e2d})), however, parameter $\xi$ starts to depend on the number of fermions \cite{Zyl}. The dependence is rather weak and for the number of $^{6}$Li atoms considered by us here it is fairly to put $\xi \approx 0.04$.  The total energy of a finite Bose-Fermi droplet is then given by $E[n_B(\mathbf{r}),n_F(\mathbf{r})] = \int d\mathbf{r} (\varepsilon_{2D} + \varepsilon_k^B + \varepsilon_{k,W}^F)$. Both of the system's constituents, bosonic and fermionic clouds, can be regarded as fluids that can be treated within standard quantum hydrodynamics \cite{Madelung,Wheeler} by introducing  the density and the velocity fields. Hydrodynamic equations can be reworked into a form of coupled set of Schr{\"o}dinger-like equations via inverse Madelung transformation \cite{Dey98,Domps98,Grochowski17}. In order to find the density profiles of the bosonic and fermionic species, the coupled set of Schr{\"o}dinger equation of motion are solved by employing imaginary time propagation technique \cite{Gawryluk17}. See {\ref{hydroapproach}} and \ref{Madelungapproach} for details.

Figure~\ref{fig3} shows the ground state densities for $^{133}$Cs$\;$-$^{6}$Li mixture for three different numbers of bosons and fermions. With increasing number of particles, as the droplets grow in size, the surface effects diminish as expected, and consequently, the peak densities approach the ones predicted by the analysis based on a uniform mixture in thermodynamic limit. The stability of the droplets is further verified by the real time propagation of the coupled set of Schr{\"o}dinger equation of motion.

\section{1D mixture}
\label{1dmix}
We now look at the possibilities for droplet formation in the case of attractive inter- and repulsive intraspecies interactions of a 1D Bose-Fermi mixture. The mean-field energy density of 1D uniform mixture is given by
\begin{equation}
\label{energyuniform}
\varepsilon_{MF}^{1d} = \frac{\hbar^2 \pi^2 n_F^3}{6 m_F}  + \frac{1}{2} g_{BB} n_B^2 + g_{BF} n_B n_F.
\end{equation}
The 1D coupling constant can be re-expressed in terms of the scattering length $g_{\sigma \sigma'}^{1d}=-\hbar^2/\left(\mu_{\sigma \sigma'} a_{\sigma \sigma'}^{1d}\right)$. The weakly interacting regime requires $|a_{\sigma \sigma'}^{1d}|n_{\sigma} \gg 1$.

The 1D LHY correction for the Bose-Fermi mixture can be obtained from Eq.~(2). Unlike the 2D case we no longer require to introduce a cut-off.  The integration over the entire momentum range can be performed analytically. We find $\varepsilon_{LHY}^{1d} = -{2 g_{BB} n_B (m_B g_{BB} n_B)^{1/2}}/{(3 \pi \hbar)}$.
The correction to the Bose-Fermi interaction in 1D,  as it can be followed from Eq.~(\ref{bf-correction}), turns out to be $\delta \varepsilon_{BF}^{1d} = -({m_B g_{BF}^2 n_B}\mathcal{I}_{1d})/({2 \pi^2 \hbar^2})$,
where
\begin{equation}
\nonumber
\mathcal{I}_{1d}= \int_{-\infty}^{\infty} dk \int_{-1}^1 dq
\frac{\left(1-\Theta \left(1-q-k\right)\right)}{\sqrt{k^2+\alpha} \left( \sqrt{k^2+\alpha} + w k + 2 w q\right)}.
\end{equation}
Finally, combining all the contributions, the  total energy of the 1D system
\begin{equation}
\varepsilon_{1D}=\varepsilon_{MF}^{1d}+\varepsilon_{LHY}^{1d}+\delta \varepsilon_{BF}^{1d}.
\end{equation}

\begin{figure}[t]
\begin{center}
\includegraphics[width=6.6cm,height=4.5cm]{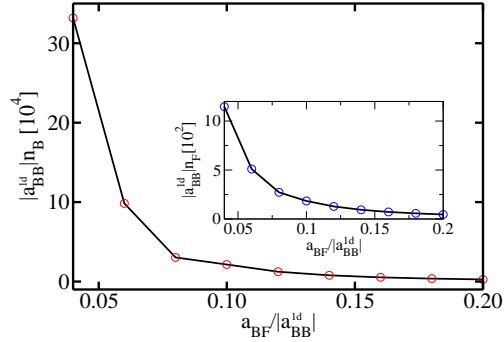}
\caption{The equilibrium density of the bosonic species in a 1D $^{133}$Cs-$^{6}$Li mixture as a function of $a_{BF}^{1d}/|a_{BB}^{1d}|$. The equilibrium density for the corresponding fermionic species is shown in inset.}
\label{fig4}
\end{center}
\end{figure}
The investigation for the possibility of a stable droplet formation in $^{133}$Cs-$^{6}$Li follows the same route discussed already in context of the 2D systems. The red circles in Fig.~\ref{fig4} show the equilibrium densities of $^{133}$Cs as function of $a_{BF}^{1d}/|a_{BB}^{1d}|$. The equilibrium densities of corresponding $^{6}$Li are shown by blue circles in the inset of Fig.~\ref{fig4}. One can see that the smaller the value  $a_{BF}^{1d}/|a_{BB}^{1d}|$, the greater the quantity $|a_{\sigma \sigma'}^{1d}|n_{\sigma}$, which must be much larger than unity in order to ensure the applicability of our theory.

It can be numerically checked that $\mathcal{I}_{1d}$ is always positive, which makes the Bose-Fermi correction term attractive in nature, similar to the LHY term. Interestingly, we find that an effectively attractive mixture satisfying all condition of droplet formation exist within the mean-field description itself. In contrast to the 2D and 3D Bose-Fermi droplets and also to the Bose-Bose droplets, the tiny correction terms, although present, do not play any crucial role in  formation of a stable droplet in 1D.

\begin{figure}[t]
\begin{center}
\includegraphics[width=7.0cm,height=4.5cm]{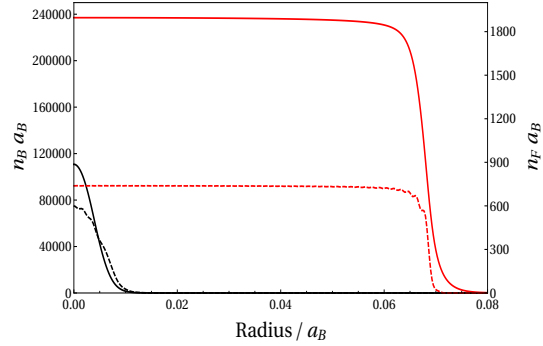}
\caption{The densities of bosonic and fermionic species in a one-dimensional $^{133}$Cs-$^{6}$Li droplet for $a_{BF}^{1d}/|a_{BB}^{1d}|=0.032$. For smaller number of atoms (black solid and dashed lines, left corner) densities resemble the solitonic solution, here $N_B=1000$ and $N_F=7$. For larger samples, here with $N_B=32000$ and $N_F=100$, a flat part in the bulk is clearly developed (red solid and dashed lines). }
\label{fig5}
\end{center}
\end{figure}

In Fig.~\ref{fig5} we show the bosonic and fermionic densities of a finite one-dimensional `mean-field' droplet for $a_{BF}^{1d}/|a_{BB}^{1d}|=0.032\,$ $(g_{BF}^{1d}/g_{BB}^{1d}=-360)$ and different numbers of atoms. In both cases $N_B/N_F$ is of the order of one hundred. Since the number of fermions considered is small, we use here an approach which treats fermionic atoms individually. We use the Hartree-Fock formalism. 
Assigning a single-particle orbital to each fermionic atom, $\phi^F_{j}({\bf r})$, and assuming all the bosons to occupy the same state $\phi_B({\bf r})$, we solve the following set of time-dependent Hartree-Fock equations
\begin{eqnarray}
i\hbar\frac{\partial \phi_B}{\partial t} = \left[-\frac{\hbar^2}{2 m_B}\nabla^2 + g_B\, n_B
 + g_{BF}\, n_F  \right] \phi_B  \,, \nonumber \\
i\hbar\frac{\partial \phi^F_j}{\partial t} = \left[-\frac{\hbar^2}{2 m_F}\nabla^2_j 
+ g_{BF}\, n_B \right] \phi^F_j,
\label{hartee_eqn}
\end{eqnarray}  
where $j=1,...,N_F$ and the total fermionic and bosonic densities are $n_F=\sum_j^{N_F} |\phi^F_j|^2$ and $n_B=N_B\,|\phi_B|^2$, as well as its time-independent version (see Ref.~\cite{Rakshit18} for details).    In this way we find self-consistently the single-fermion orbitals, i.e. one-particle eigenstates of the effective potential created by both bosonic and fermionic components and hence densities.    Clearly, as Fig.~\ref{fig5} suggests, going to larger samples the solution changes its character from the soliton-like (black solid and dashed lines, left corner) reported by us already some time ago (Ref. \cite{Karpiuk04}) to the one with fully developed flat part typical for droplets (red solid and dashed lines). Noticeably, appearance of a train of Bose-Fermi solitons for appropriately tuned Bose-Fermi interaction, as claimed in Ref.~\cite{Karpiuk04}, has been confirmed in recent experiment \cite{chin18}. In 1D the fermions can be considered as impurities -- much smaller in fraction in comparison to 2D and 3D \cite{Rakshit18} cases, yet essential for the droplet formation.

\section{Discussion}
\label{conclude}
In conclusion, in this work we show that higher-order quantum corrections may lead an ultracold weakly interacting gas of Bose-Fermi mixture towards a self-bound liquid state of matter. We find that in contrast to the 3D case \cite{Rakshit18}, where droplets emerge above certain critical value of $a_{BF}/a_{BB}$, 2D ultradilute liquids are formed for arbitrary intra-species and inter-species  interactions. Interestingly in 1D geometry, no quantum fluctuations are needed to form self-bound droplets.

Bose-Fermi systems are exotic due to the non-trivial interplay between dimension dependent scaling of kinetic energy of fermions, mean-field interactions and their higher-order contribution. The role of quantum pressure of fermions decreases while going from 3D to 1D geometry. Unlike in 3D systems, where fermionic kinetic energy scales like $n_F^{5/3}$, in  2D space the kinetic energy scales as $n_F^2$,
and in 1D the scaling is $n_F^3$. This is why self-bound droplets in 2D can be formed for almost vanishing  mean-field energy what is not the case for the 3D system. While recent experiments with Bose-Bose droplets have vindicated the role of beyond-mean-field LHY correction, the quantum Bose-Fermi droplets promises to be an ideal platform for probing another higher order quantum correction in Bose-Fermi interaction originating from the density fluctuations of bosonic and fermionic species. 

The quadratic scaling of kinetic energy on fermionic density in 2D makes these systems closely analogous to Bose-Bose systems and hence  the most promising candidates for investigating the liquid phases in highly anisotropic pancake shaped geometry with tight harmonic trap in the transverse direction. 
In future, it will be interesting to understand the nature of higher-order quantum correction in Bose-Fermi interaction, and the role they play in liquid formation, throughout the entire dimensional crossovers \cite{Zin18}. 

\section*{Acknowledgments}
DR and PZ acknowledge support from the EU Horizon 2020-FET QUIC 641122. MG, MB, and TK acknowledge support from the (Polish) National Science Center Grant No. 2017/25/B/ST2/01943. ML acknowledges the Spanish Ministry MINECO (National Plan 15 Grant: FISICATEAMO No. FIS2016-79508-P, SEVERO OCHOA No. SEV-2015-0522, FPI), European Social Fund, Fundaci{\'o} Cellex, Generalitat de Catalunya (AGAUR Grant No. 2017 SGR 1341 and CERCA/Program), ERC AdG OSYRIS, and EU FETPRO QUIC.

\appendix

\section{Justification for the choice of the cut-off }
\label{justification}

For a two-dimensional Bose gas alone, in order to reproduce low-energy scattering properties of the exact potential, one has \cite{Castin03,Petrov16}
\begin{equation}
\frac{1}{g_{BB}} = \frac{m_B}{2\pi \hbar^2} (\ln{2/(\kappa\, a_{BB}) - \gamma})  \,,
\label{Bgasalone}
\end{equation}
where $\kappa$ is a cut-off momentum. The formula is equivalent to the one used in the main text. Energy of a two-dimensional Bose gas is then given by
\begin{eqnarray}
E_B &=& \frac{1}{2} g_{BB}\, n_B N_B + \frac{1}{2} \sum_{k<\kappa}{\left( \sqrt{\hbar^2k^2/2m_B (\hbar^2 k^2/2m_B + 2 g_{BB} n_B)} \right. }  \nonumber  \\
&-&  \left. \hbar^2 k^2/2m_B - g_{BB} n_B \right)  
\label{Bgasenergy}
\end{eqnarray}
After integrating over momenta, the energy density becomes
\begin{equation}
\varepsilon_B = \frac{1}{2} g_{BB} n_B^2 +  \frac{(g_{BB} n_B)^2 m_B}{8 \pi \hbar^2}  \ln\left(\frac{g_{BB} n_B \sqrt{e}}{\hbar^2\kappa^2/m_B}\right) 
\label{B2denergy}
\end{equation}
and the first (mean-field) and the second (LHY correction) terms can be already identified in Eqs. (\ref{emf}) and (\ref{elhy1}), respectively, in the main text. By introducing $g_{BB}$ from (\ref{Bgasalone}) into Eq. (\ref{B2denergy}) one gets
\begin{equation}
\frac{\varepsilon_B}{n_B}\, \frac{m_B\,a_{BB}^2}{2\pi\,\hbar^2} = \frac{x}{|\ln{y}|} + \frac{x}{\ln^2\!y} \ln{\left(\frac{1}{y |\ln{y}|} \pi\, e^{2\gamma+1/2}\right)}  \,,
\label{energyxy}
\end{equation}
where $x=n_B a_{BB}^2$ and $y=(\kappa\, a_{BB}/ 2 e^{-\gamma})^2$ are dimensionless parameters. In the limit of small values of parameter $x$, the energy density is given by Schick formula \cite{Schick71}
\begin{equation}
\varepsilon_{Schick} /n_B = - \frac{\hbar^2}{m_B} \frac{2\pi\, n_B}{\ln{(n_B a_{BB}^2)}} = \frac{2\pi\, \hbar^2}{m_B\, a_{BB}^2} \frac{x}{|\ln{x}|}  \,.
\label{eSchick}
\end{equation}
Therefore
\begin{equation}
\varepsilon_B / \varepsilon_{Schick} = \frac{|\ln{x}|}{|\ln{x z}|} + \frac{|\ln{x}|}{\ln^2\!(x z)} \ln{\left(\frac{1}{z |\ln{x z}|} \pi\, e^{2\gamma+1/2}\right)}  \,,
\label{eSchickratio}
\end{equation}
where a new, density dependent, cut-off $z=y/x \propto \kappa^2/n_B$ has been introduced. Obviously, the second term in (\ref{eSchickratio}) vanishes in the limit of small $x$, whereas the first one approaches one. The low density limit for the Bose gas is then recovered. The energy becomes flat in a wider range of the cut-off parameter $z$ when $x$ gets smaller.

However, in this paper we consider a two-dimensional Bose-Fermi mixture and description of boson-fermion interaction in two dimensions also requires introducing an appropriate cut-off momentum. We equalize both cut-offs and propose an iterative procedure to find its value. We start with, in principle, arbitrary chosen initial value of the cut-off but the value obtained by the minimum stability condition, Eq. (\ref{cut-off}), could be a good guess. Then the bosonic and fermionic densities are obtained based on Eqs. (\ref{zero-pressure}) and (\ref{zero-pressure-2}). Next, we utilize the condition $\partial \varepsilon_{2D}/\partial \kappa^2 =0$ to get a new value of $\kappa$. This is because the total energy of the system should only weakly depend on the cut-off. Therefore, the value of the cut-off should be close to the position of the wide maximum of the total energy, $\varepsilon_{2D}(\kappa)$ for fixed densities. Then the next iterations are performed until the bosonic and fermionic densities, as well as position of maximum do not change. The value of the cut-off found in this way is $\kappa_c=0.0340$ and the densities are: $n_B=3.97\times 10^{-5}\,a_{BB}^{-2}$ and $n_F=4.10\times 10^{-6}\,a_{BB}^{-2}$, see Fig. \ref{fig1}.

\begin{figure}[htb]
\begin{center}
\includegraphics[width=6.0cm,height=4.5cm]{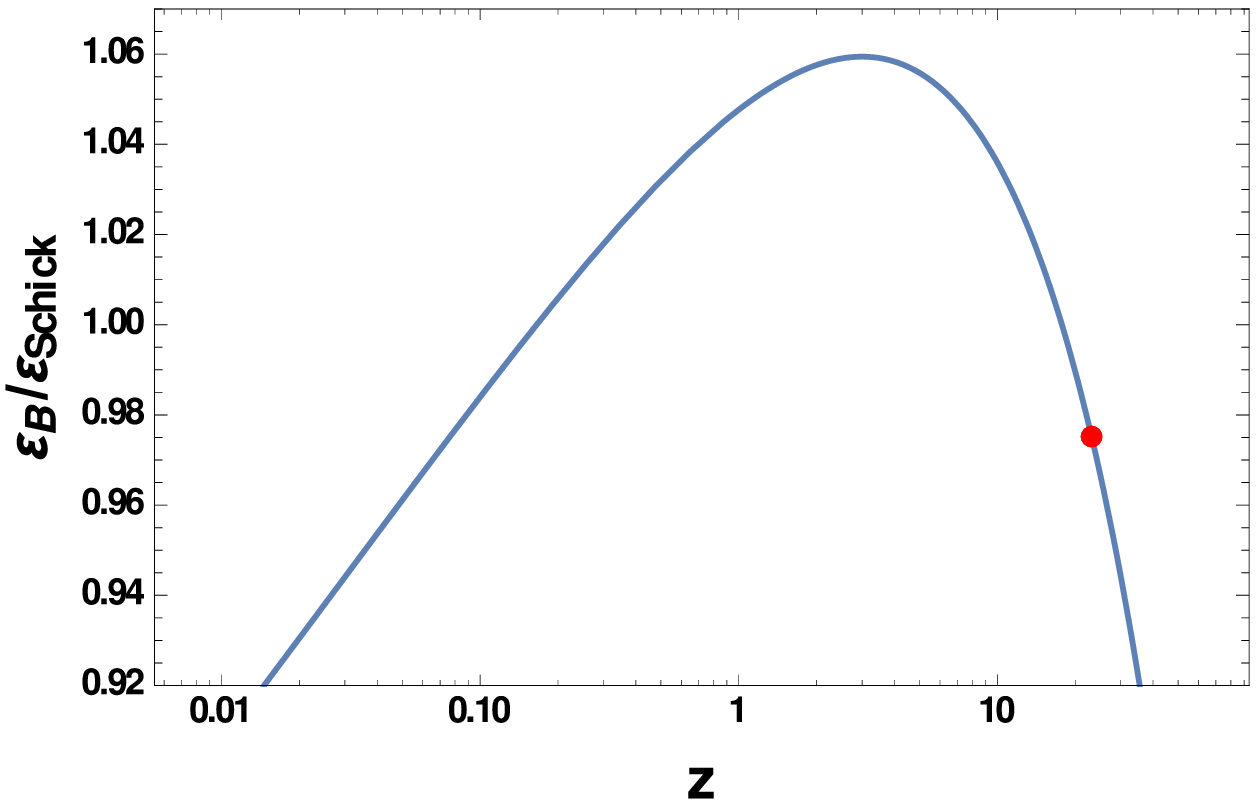}  \includegraphics[width=6.0cm,height=4.5cm]{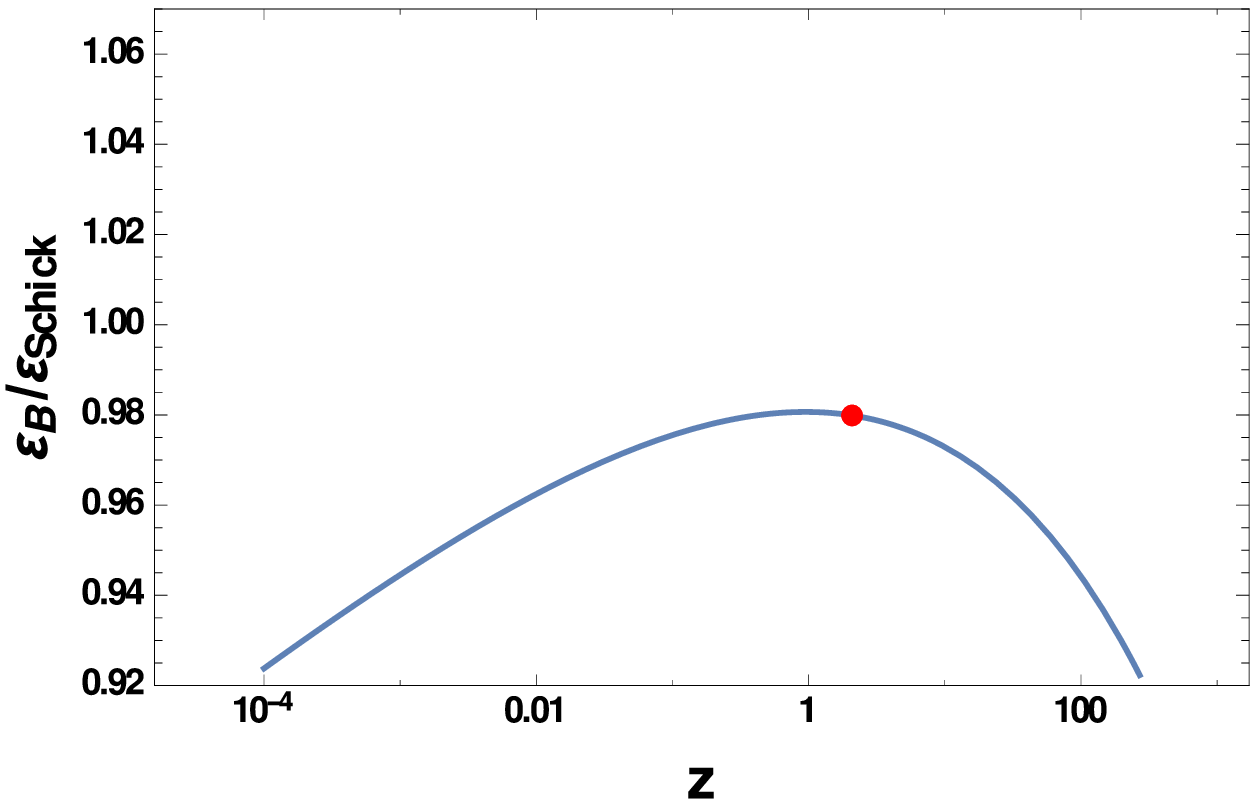}
\caption{Energy of bosonic component of a two-dimensional Bose-Fermi mixture as a function of density dependent cut-off momentum $z$. Left frame: $a_{BF}=10^4 a_{BB}$ and $n_B a_{BB}^2=3.97\times 10^{-5}$. Right frame: $a_{BF}=10^{10} a_{BB}$ and $n_B a_{BB}^2=4.07\times 10^{-13}$. The blue solid line is the formula (\ref{eSchickratio}) whereas red dots show the value of energy density at the cut-off momentum found via iterative procedure.  }
\label{figA1}
\end{center}
\end{figure}

We expect that even after adding fermions to bosons, when mutual attraction becomes infinitely small, $n_B a_{BF}^2 \rightarrow \infty$, the energy of bosonic component will still satisfy the limit of low densities, the Schick formula. It is indeed the case, as shown in Fig. \ref{figA1}. The bosonic energy becomes more and more flat when the bosonic density gets smaller and the value of  cut-off approaches the position corresponding to the maximum of $\varepsilon_B / \varepsilon_{Schick}$, which in turn approaches the value of one as expected (red dot in Fig. \ref{figA1}). This way we show that the cut-off we choose allows to recover the standard expression of energy of weakly interacting Bose gas in 2D.

\section{Hydrodynamic equations}
\label{hydroapproach}
Hydrodynamic equations for a gas of neutral fermionic or bosonic atoms can be derived based on quantum kinetic equations for reduced density matrices \cite{Frolich,Wong,MarchDeb}. From the whole hierarchy of equations, the one for the one-particle density matrix is of most practical interest
\begin{eqnarray}
i\hbar \frac{\partial}{\partial t} \rho_1(\vec{r}_1,\vec{r}_2,t) = -\frac{
\hbar^2}{2 m} (\vec{\nabla}_1^2 - \vec{\nabla}_2^2) \, \rho_1(\vec{r}_1,\vec{
r}_2,t) \phantom{11111} & &  \nonumber \\
+ \int d^3 r^{\prime} \left[V(\vec{r}_1-\vec{r}^{\,\prime}) - V(\vec{r}_2-
\vec{r}^{\,\prime})\right] \rho_2(\vec{r}_1,\vec{r}^{\,\prime};\vec{r}_2,
\vec{r}^{\,\prime},t) & &  \nonumber \\
+ \left[V_{ext}(\vec{r}_1,t)-V_{ext}(\vec{r}_2,t)\right] \rho_1(\vec{r}_1,
\vec{r}_2,t) \;, \phantom{11111111111l} & &  \label{densitymat}
\label{red1}
\end{eqnarray}
where $V(\vec{r}_1-\vec{r}_2)$ is the two-particle interaction term and $V_{ext}(\vec{r},t)$ is the external potential. Eq. (\ref{red1}) involves the two-particle density matrix $\rho_2$. In the limit $\vec{r}_1 \rightarrow \vec{r}_2$ ($=\vec{r}$), Eq. (\ref{densitymat}) results in the continuity equation 
\begin{eqnarray}
\frac{\partial n(\vec{r},t)}{\partial t} + \vec{\nabla} \cdot \left[n(\vec{r},t) \, \vec{v}(\vec{r},t)\right] & = & 0 \;,  
\label{hydr1}
\end{eqnarray}
where the density and velocity fields are defined as follows:
\begin{eqnarray}
n(\vec{r},t) & = & \lim_{\vec{r}_1 \rightarrow \vec{r}_2} \rho_1(\vec{r}
_1,\vec{r}_2,t)  \nonumber \\
\vec{v}(\vec{r},t) & = & \frac{\hbar}{2 m} \lim_{\vec{r}_1 \rightarrow \vec{r
}_2} (\vec{\nabla}_1 - \vec{\nabla}_2) \, \chi(\vec{r}_1,\vec{r}_2,t)
\label{deffield}
\end{eqnarray}
and $\chi(\vec{r}_1,\vec{r}_2,t)$ is the phase of the one-particle density matrix $\rho_1(\vec{r}_1,\vec{r}_2,t)=\sigma(\vec{r}_1,\vec{r}_2,t) \exp{[i\chi(\vec{r}_1,\vec{r}_2,t)]}$.

Eq. (\ref{densitymat}) can be rewritten by introducing the center-of-mass, $\vec{r}=(\vec{r}_1+\vec{r}_2)/2$, and the relative position, $\vec{s}=\vec{r}_1-\vec{r}_2$, coordinates. Then, by taking the derivative of Eq. (\ref{densitymat}) with respect to the coordinate $\vec{s}$, the hydrodynamic Euler-type equation of motion is obtained in the limit $\vec{s} \rightarrow 0$ 
\begin{eqnarray}
\frac{\partial \vec{v}(\vec{r},t)}{\partial t} & = & -\frac{\vec{\nabla}
\cdot \mathrm{T}}{m\, n(\vec{r},t)} - [\vec{v}(\vec{r},t) \cdot \vec{
\nabla}] \, \vec{v}(\vec{r},t)  \nonumber \\
& + & \frac{\vec{F}(\vec{r},t)}{m\, n(\vec{r},t)} -\frac{\vec{\nabla
} V_{ext}(\vec{r},t)}{m} \;,  \label{hydr2}
\end{eqnarray}
where the kinetic-energy stress tensor, $\mathrm{T}$, is given by 
\begin{eqnarray}
\mathrm{T}_{kl} & = & -\frac{\hbar^2}{m} \lim_{\vec{s} \rightarrow 0} \frac{
\partial^{\,2} \sigma(\vec{r},\vec{s},t)} {\partial s_k \partial s_l} 
\label{kin}
\end{eqnarray}
and the force, $\vec{F}(\vec{r},t)$, resulting from interactions between atoms is
\begin{eqnarray*}
\vec{F}(\vec{r},t) & = & -\int d^3 r^{\prime} \, \vec{\nabla}_{\vec{r}
} \, V(\vec{r}-\vec{r}^{\,\prime}) \, \rho_2(\vec{r},\vec{r}^{\,\prime};\vec{
r},\vec{r}^{\,\prime},t) \;.
\end{eqnarray*}

The kinetic-energy stress tensor for fermions can be easily calculated within the Thomas-Fermi approximation. The one-particle Wigner function within the Thomas-Fermi approximation in a three-dimensional space is given by 
\begin{eqnarray*}
w(\vec{r}, \vec{p}) & = & \eta(\hbar^2[6\pi^2 n_F(\vec{r})] ^{2/3} - \vec{p}
^{\; 2}) \;,
\end{eqnarray*}
where $\eta()$ is the unit step function. The one-particle density matrix is calculated according to 
\begin{eqnarray*}
\rho_1(\vec{r}, \vec{s}) & = & \int \frac{d^3 p}{(2\pi \hbar)^3} w(\vec{r}, 
\vec{p}) e^{i \vec{p} \vec{s} / \hbar} \;,
\end{eqnarray*}
which implies that
\begin{eqnarray*}
\rho_1(\vec{r}, \vec{s}) & = & \frac{2}{(2\pi \hbar)^2} \frac{1}{s} \left[ - 
\frac{\hbar p_F}{s} \cos \left(\frac{p_F s}{\hbar} \right) + \frac{\hbar^2}{
s^2} \sin \left(\frac{p_F s}{\hbar} \right) \right] ,
\end{eqnarray*}
where $p_F(\vec{r})=\hbar\, (6\pi^2 n_F(\vec{r}))^{1/3}$ is the local Fermi momentum. Then, from Eq. (\ref{kin}) one gets the kinetic-energy stress tensor $\mathrm{T}_{kl}=[\hbar^2 /(30 \pi^2 m_F)] (6 \pi^2 n_F)^{5/3}\, \delta_{kl}$. Assuming the case of non-interacting fermions (spin-polarized atoms at low temperature), the Eqs. (\ref{hydr1}) and (\ref{hydr2}) become a closed set of hydrodynamic equations
\begin{eqnarray}
&& \frac{\partial n_F}{\partial t} + \vec{\nabla} \cdot \left( n_F \vec{v}_F \right) = 0 \nonumber \\
&& \frac{\partial \vec{v}_F}{\partial t} + (\vec{v}_F \cdot \vec{\nabla}) \, \vec{v}_F
+ \vec{\nabla} \left( \frac{\hbar^2}{2 m_F^2} (6 \pi^2)^{2/3}\; n_F^{2/3} + \frac{V_{\mathrm{ext}}}{m_F} \right) = 0 \;. 
\nonumber \\
&&  \nonumber \\
\label{hydrofer}
\end{eqnarray}

There is, of course, a space for improvements here. Commonly, gradient corrections are locally added to the Thomas-Fermi kinetic energy density (diagonal part of the kinetic-energy stress tensor), the simplest one being the Weizs\"acker correction in the form of $\mathrm{T}_W=\xi (\hbar^2/8 m_F)(\vec{\nabla} n_F)^2/n_F$, with $\xi=1/9$ \cite{Weizsacker,Kirznits,Oliver}. Then the last term in  the second equation of Eqs. (\ref{hydrofer}) is modified by adding the derivative $\delta \mathrm{T}_W / \delta n_F$.

We further assume that the fermionic flow is irrotational, i.e. the vorticity vanishes, $\vec{\nabla} \times \vec{v}_F =0$. By using one of the vector identities, $(\vec{v}_F \cdot \vec{\nabla}) \, \vec{v}_F=(\vec{\nabla} \times \vec{v}_F) \times \vec{v}_F + \vec{\nabla}(\vec{v}_F^{\:2}) /2$, Eqs. (\ref{hydrofer}) are turned into the following ones
\begin{eqnarray}
&& \frac{\partial n_F}{\partial t} = - \vec{\nabla} \cdot \left( n_F \vec{v}_F \right)  \nonumber \\
&& m_F \frac{\partial \vec{v}_F}{\partial t} = - \vec{\nabla} \left( \frac{\hbar^2}{2 m_F} (6 \pi^2)^{2/3}\; n_F^{2/3} + 
\frac{\delta \mathrm{T}_W}{\delta n_F} + \frac{m_F \vec{v}_F^{\:2}}{2} + V_{\mathrm{ext}} \right) \;. 
\nonumber \\
&&  \nonumber \\
\label{hydrofernew}
\end{eqnarray}
The set of equations (\ref{hydrofer}) was first used by Ball {\it et al.} \cite{Wheeler} to study the oscillations of electrons in a many-electron atom induced by ultraviolet and soft x-ray photons.

For a gas of neutral bosonic atoms, in the mean-field approximation when the system's many-body wave function is $\varphi(\vec{r}_1)\,\varphi(\vec{r}_2) \cdot...\cdot \varphi(\vec{r}_N)$, the one-particle density matrix is given by
\begin{eqnarray*}
\rho_1(\vec{r}, \vec{s}) & = & \varphi(\vec{r}+\frac{1}{2} \vec{s}) \;
\varphi^*(\vec{r}-\frac{1}{2} \vec{s}) \;.
\end{eqnarray*}
The kinetic-energy stress tensor possesses now off-diagonal elements and leads to the so-called quantum pressure term in the equation of motion. Assuming only contact interactions between bosonic atoms (with $a$ being the scattering length), the Eqs. (\ref{hydr1}) and (\ref{hydr2}) become the set of the following equations
\begin{eqnarray}
&& \frac{\partial n_B}{\partial t} = - \vec{\nabla} \cdot \left( n_B \vec{v}_B \right)  \nonumber \\
&& m_B \frac{\partial \vec{v}_B}{\partial t} = - \vec{\nabla} \left( \frac{4 \pi \hbar^{2} a_B}{m_B} n_B + m_B \frac{\vec{v}_B^{\:2}}{2} + V_{\mathrm{ext}} - \frac{\hbar^{2}}{2 m_B} \frac{\vec{\nabla}^{2} \sqrt{n_B}}{\sqrt{n_B}} \right) \;.
\nonumber \\
&&  \nonumber \\
\label{hydrobos}
\end{eqnarray}
Eqs. (\ref{hydrobos}) are, in fact, the hydrodynamic representation of the Gross-Pitaevskii equation. They can be improved with respect to the boson-boson interactions by considering the beyond mean-field correction -- the Lee-Huang-Yang correction.

The mixture of bosonic and fermionic atoms is then described by equations (\ref{hydrofernew}) and (\ref{hydrobos}) modified by a term representing the interactions between bosons and fermions. This term, of the form of $\delta E_{BF} / \delta n_F$ for fermions and $\delta E_{BF} / \delta n_B$ for bosons, where $E_{BF}$ is the total Bose-Fermi interaction energy, acts as an additional potential in the Euler-like equations of motion. What is crucial for the existence of Bose-Fermi droplets, $E_{BF}$ energy includes the beyond mean-field correction. Although the above derivation of hydrodynamics of the Bose-Fermi mixture was performed explicitly in a three-dimensional space, it can be repeated in two- and one-dimensional cases as well. The changes which should be done are related to the local fermionic kinetic energy (for instance, in a two-dimensional space $\mathrm{T}_{kl}=(\hbar^2 \pi / m_F)\, n_F^2\, \delta_{kl}$ and $\mathrm{T}_W=\xi (\hbar^2/8 m_F)(\vec{\nabla} n_F)^2/n_F$, with $\xi=0.04$) and beyond mean-field corrections for Bose-Bose and Bose-Fermi interaction terms.

\section{Inverse Madelung transformation}
\label{Madelungapproach}
As shown in the previous Section, the Bose-Fermi mixture can be treated by using equations like (\ref{hydrofernew}) and (\ref{hydrobos}). The very convenient way to further tackle Eqs. (\ref{hydrofernew}) and (\ref{hydrobos}) is to put them in a form of the Schr\"odinger-like equations by using the inverse Madelung transformation \cite{Madelung}. This is just a mathematical transformation which introduces the single complex function instead of density and velocity fields used in a hydrodynamic description. Both treatments are equivalent provided the velocity field is irrotational (vanishing vorticity). This assumption is obviously fullfilled for bosonic component since we consider that bosons populate a single quantum state. It is also true for fermions in a stationary case (zero velocity field) and in the case of dynamics studied by us while the trap confining mixture is adiabatically removed (in this case we a posteriori check that vorticity is zero).

After the inverse Madelung transformation is applied:
\begin{equation}
{\partial \psi_F}(\vec{r},t)=\sqrt{n_F(\vec{r},t)}e^{-i \chi(\vec{r},t)}
\end{equation}
where $\vec{v}(\vec{r},t)=\vec{\nabla}\chi(\vec{r},t)$, Eqs. (\ref{hydrofernew}) and (\ref{hydrobos}) are turned into
\begin{eqnarray}
i\hbar\frac{\partial \psi_F}{\partial t} &=& \left[-\frac{\hbar^2}{2 m_F}\nabla^2 +  \frac{\xi' \hbar^2}{2 m_F} \frac{\nabla^2 |\psi_F|}{\psi_F} + \beta |\psi_F|^2 + g_{BF}\, |\psi_B|^2 \right. \nonumber \\   
&+& \left. \mathcal{C} \,  \mathcal{I}_c (\omega,\alpha)  + \mathcal{C} \,|\psi_F|^2 \frac{\partial \mathcal{I}_c}{\partial \alpha} \frac{\partial \alpha}{\partial n_F}\right] \psi_F , 
\nonumber \\ 
i\hbar\frac{\partial \psi_B}{\partial t} &=& \left[-\frac{\hbar^2}{2 m_B}\nabla^2 + \beta |\psi_B|^2 + g_{BF}\, |\psi_F|^2 +\mathcal{A} |\psi_B|^2 \right. 
\nonumber \\   
&+& \left. 2 \mathcal{A} |\psi_B|^2 \ln(\mathcal{B} |\psi_B|^2) + \mathcal{C} \,|\psi_F|^2 \frac{\partial \mathcal{I}_c}{\partial \alpha} \frac{\partial \alpha}{\partial n_B} \right] \psi_B,
\label{2D}
\end{eqnarray}
where now the main objects to be solved are the bosonic wave function, $\psi_B$, and the fermionic pseudo-function, $\psi_F$. The coefficients are: $\xi'=1-\xi=0.96$, $\mathcal{A}=\frac{g_{BB}^2}{8 \pi \hbar^2}$, $\mathcal{B}=\frac{g_{BB} \sqrt{e}}{(\hbar^2 \kappa^2/m_B)}$, and $\mathcal{C}=-\frac{(4 \pi)^2 m_B g_{BF}^2}{(2 \pi)^4 \hbar^2}$. The bosonic wave function and the fermionic pseudo-wave function are normalized as $N_{B,F} = \int d\mathbf{r}\, |\psi_{B,F}|^2$. The integral $\mathcal{I}_c$ and all the parameters $\omega$, $\alpha$, and $\beta$ and those appearing in coefficients $\mathcal{A}$, $\mathcal{B}$, and $\mathcal{C}$     are defined in the main text.

Note that Eqs. (\ref{2D}) describe two-dimensional Bose-Fermi mixture with beyond mean-field Lee-Huang-Yang correction for boson-boson interactions and analogous correction for the interaction between bosons and fermions.


\section*{References}
\begin{thebibliography}{10}

\bibitem{Kadau16} H. Kadau, M. Schmitt, M. Wenzel, C. Wink, T. Maier, I. Ferrier-Barbut, and T. Pfau, \emph{Observing the Rosensweig instability of a quantum ferrofluid}, Nature {\bf 530}, 194 (2016).
\bibitem{Ferrier16a} I. Ferrier-Barbut, H. Kadau, M. Schmitt, M. Wenzel, and T. Pfau, \emph{Observation of quantum droplets in a strongly dipolar Bose gas}, Phys. Rev. Lett. {\bf 116}, 215301 (2016).
\bibitem{Schmitt16} M. Schmitt, M. Wenzel, B. B\"ottcher, I. Ferrier-Barbut, and T. Pfau, \emph{Self-bound droplets of a dilute magnetic quantum liquid}, Nature {\bf 539}, 259 (2016).
\bibitem{Chomaz16} L. Chomaz, S. Baier, D. Petter, M.J. Mark, F. W\"achtler, L. Santos, and F. Ferlaino, \emph{Quantum-fluctuation-driven crossover from a dilute Bose-Einstein condensate to a macrodroplet in a dipolar quantum fluid}, Phys. Rev. X {\bf 6}, 041039 (2016).
\bibitem{Cabrera17} C.R. Cabrera, L. Tanzi, J. Sanz, B. Naylor, P. Thomas, P. Cheiney, and L. Tarruell, \emph{Quantum liquid droplets in a mixture of Bose Einstein condensates}, Science {\bf 359}, 301 (2018).
\bibitem{Tarruell17b} P. Cheiney, C. R. Cabrera, J. Sanz, B. Naylor, L. Tanzi, L. Tarruell, \emph{Bright soliton to quantum droplet transition in a mixture of Bose-Einstein condensates}, Phys. Rev. Lett. {\bf 120}, 135301 (2018).
\bibitem{Fattori17} G. Semeghini, G. Ferioli, L. Masi, C. Mazzinghi, L. Wolswijk, F. Minardi, M. Modugno, G. Modugno, M. Inguscio, M. Fattori,  \emph{Self-bound quantum droplets of atomic mixtures in free space}, Phys. Rev. Lett. {\bf 120}, 235301 (2018).

\bibitem{Petrov15} D. S. Petrov, \emph{Quantum mechanical stabilization of a collapsing Bose-Bose mixture}, Phys. Rev. Lett. {\bf 115}, 155302 (2015).
\bibitem{Wachtler16a} F. W\"achtler and L. Santos, Quantum filaments in dipolar Bose-Einstein condensates, Phys. Rev. A {\bf 93}, 061603(R) (2016).
\bibitem{Baillie16} D. Baillie, R. M. Wilson, R. N. Bisset, and P. B. Blakie, \emph{Self-bound dipolar droplet: A localized matter wave in free space}, Phys. Rev. A {\bf 94}, 021602(R) (2016).
\bibitem{Rafal16} R. O{\l}dziejewski and K. Jachymski, \emph{Properties of strongly dipolar Bose gases beyond the Born approximation}, Phys. Rev. A {\bf 94}, 063638 (2016).
\bibitem{Cui18} X. Cui, \emph{Spin-orbit coupling induced quantum droplet in ultracold Bose-Fermi mixtures}, Phys. Rev. A {\bf 98}, 023630 (2018).

\bibitem{Rakshit18} D. Rakshit, T. Karpiuk, M. Brewczyk and M. Gajda, \emph{Quantum Bose-Fermi droplets}, arXiv:1801.00346.

\bibitem{Petrov16} D. S. Petrov and G. Astrakharchik, \emph{Ultradilute low-dimensional liquids}, Phys. Rev. Lett. {\bf 117}, 100401 (2016).
\bibitem{Mishra16} C. Mishra, D. Edler, F. W\"achtler, R. Nath, S. Sinha, and L. Santos, \emph{Quantum droplets in one-dimensional dipolar Bose-Einstein condensates}, arXiv:1610.09176 (2016).
\bibitem{Zin18} P. Zin, M. Pylak, T. Wasak, M. Gajda and Z. Idziaszek, \emph{Quantum Bose-Bose droplets at a dimensional crossover}, arXiv:1805.11186.
\bibitem{Ilg18} T. Ilg, J. Kumlin, L. Santos, D. S. Petrov, H. P. B{\"u}chler, \emph{Dimensional crossover for the beyond-mean-field correction in Bose gases}, arXiv:1806.01784.

\bibitem{Popov71} V. N. Popov, \emph{To the theory of superfluidity of the twodimensional and one-dimensional Bose systems}, Teor. Mat. Fiz. {\bf 11}, 354 (1971), [Theor. Math. Phys. {\bf 11}, 565 (1972)].

\bibitem{Castin03} C. Mora and Y. Castin, \emph{Extension of Bogoliubov theory to quasicondensates}, Phys. Rev. A 67, 053615 (2003).

\bibitem{Lee57} T. D. Lee, K. Huang, and C. N. Yang, \emph{Eigenvalues and eigenfunctions of a Bose system of hard spheres and its low-temperature properties}. Phys. Rev. {\bf 106}, 1135 (1957).

\bibitem{Wilkens02} A.P. Albus, S.A. Gardiner, F. Illuminati, and M. Wilkens, Quantum field theory of dilute homogeneous Bose-Fermi mixtures at zero temperature: General formalism and beyond mean-field corrections, Phys. Rev. A {\bf 65}, 053607 (2002).
\bibitem{Viverit02} L. Viverit and S. Giorgini, Ground-state properties of a dilute Bose-Fermi mixture, Phys. Rev. A 66, 063604 (2002).
\bibitem{Zhai11} Z.-Q. Yu, S. Zhang, and H. Zhai, Stability condition of a strongly interacting boson-fermion mixture across an interspecies Feshbach resonance, Phys. Rev. A {\bf 83}, 041603(R) (2011).

\bibitem{Schick71} M. Schick, \emph{Two dimensional system of hard-core bosons}, Phys. Rev. A {\bf 3}, 1067 (1971).


\bibitem{monte} The universality of such treatment adopted for 2D system can be surmised from the work in \cite{Petrov16}, which treats the Bose-Bose system on an equivalent ground, but further back it up by evidences supported by Monte-Carlo analysis. 


%







\bibitem{Weizsacker} C. F. Weizs\"acker, Z. Phys. {\bf 96}, 431 (1935).
\bibitem{Kirznits} D. A. Kirznits, Sov. Phys. JETP {\bf 5}, 64 (1957).
\bibitem{Oliver} G. L. Oliver and J. P. Perdew, \emph{Spin-density gradient expansion for the kinetic energy}, Phys. Rev. A {\bf 20}, 397 (1979).

\bibitem{Zyl} B. P. van Zyl, E. Zaremba, and P. Pisarski, \emph{Thomas-Fermi-von Weizs{\"a}cker theory for a harmonically trapped, two-dimensional, spin-polarized dipolar Fermi gas}, Phys. Rev. A {\bf 87}, 043614 (2013).


\bibitem{Madelung} E. Madelung, Z. Phys. {\bf 40}, 322 (1927).
\bibitem{Wheeler} J. A. Ball, J. A. Wheeler, and E. L. Fireman, \emph{Photoabsorption and charge oscillation of the Thomas-Fermi atom}, Rev. Mod. Phys. {\bf 45}, 333 (1973).

\bibitem{Dey98} B. Kr. Dey and B. M. Deb, \emph{Femtosecond quantum fluid dynamics of helium atom under an intense laser field}, Int. J. Quantum Chem. {\bf 70}, 441 (1998).
\bibitem{Domps98} A. Domps, P.-G. Reinhard, and E. Suraud, \emph{Time-dependent Thomas-Fermi approach for electron dynamics in metal clusters}, Phys. Rev. Lett. {\bf 80}, 5520 (1998).
\bibitem{Grochowski17} P. T. Grochowski, T. Karpiuk, M. Brewczyk, and K. Rz\c{a}\.zewski, \emph{Unified description of dynamics of a repulsive two-component Fermi gas}, Phys. Rev. Lett. {\bf 119}, 215303 (2017).
%

\bibitem{Gawryluk17} K. Gawryluk, T. Karpiuk, M. Gajda, K. Rz\c{a}{\.z}ewski, and M. Brewczyk, \emph{Unified way for computing dynamics of Bose-Einstein condensates and degenerate Fermi gases}, Int. J. Comput. Math. DOI: 10.1080/00207160.2017.1370545 (2017).

\bibitem{Karpiuk04} T. Karpiuk, M. Brewczyk, S. Ospelkaus-Schwarzer, K. Bongs, M. Gajda, and K. Rz\c{a}\.zewski, \emph{Soliton trains in Bose-Fermi mixtures}, Phys. Rev. Lett. {\bf 93}, 100401 (2004).
\bibitem{chin18} B. J. DeSalvo, K. Patel, G. Cai, and C. Chin, Fermion-Mediated Interactions Between Bosonic Atoms, arXiv:1808.07856.

\bibitem{Frolich}  H. Fr\"{o}lich, Physica {\bf 37}, 215 (1967).
\bibitem{Wong} C.Y. Wong and J.A. McDonald, \emph{Dynamics of nuclear fluid. III. General considerations on the kinetic theory of quantum fluids}, Phys. Rev. C {\bf 16}, 1196 (1977).
\bibitem{MarchDeb} N.H. March and B.M. Deb, {\it The single-particle density in physics and chemistry} (Academic Press, London, 1987).



%
%
%
%
%
%
%
%
%
%
%
%
%
%
%
%
%
%
%
%
%
%
%
%
%
%
%
%




\end{thebibliography}
 \end{document}